\documentclass[fleqn,usenatbib]{mnras}

\usepackage{newtxtext,newtxmath}

\usepackage[T1]{fontenc}

\DeclareRobustCommand{\VAN}[3]{#2}
\let\VANthebibliography\thebibliography
\def\thebibliography{\DeclareRobustCommand{\VAN}[3]{##3}\VANthebibliography}

\usepackage{graphicx}	
\usepackage{amsmath}	

\usepackage{amssymb}	

\title[A spectral hardening of 1ES 0502+675]{A spectral hardening in the Fermi-LAT Data of 1ES 0502+675}

\author[Zeng et al.]{
Yuhang Zeng,$^{1,2}$
Dahai Yan,$^{1}$\thanks{E-mail: yandahai@ynao.ac.cn}
Wen Hu,$^{3}$
and Jiancheng Wang$^{1}$
\\
$^{1}$Key Laboratory for the Structure and Evolution of Celestial Objects, Yunnan Observatories, Chinese Academy of Sciences,\\
Kunming 650011, People's Republic of China\\
$^{2}$University of Chinese Academy of Science, Beĳing 100049, People’s Republic of China\\
$^{3}$Department of Physics, Jinggangshan University, Jiangxi Province, Ji'an 343009, People's Republic of China
}

\date{Accepted XXX. Received YYY; in original form ZZZ}

\pubyear{20XX}

\begin{document}
\label{firstpage}
\pagerange{\pageref{firstpage}--\pageref{lastpage}}
\maketitle

\begin{abstract}
The $\gamma$-ray spectral feature of the blazar 1ES 0502+675 is investigated by using Fermi
Large Area Telescope (Fermi-LAT) Pass 8 data (between 100 MeV and 300 GeV) covering from 2008 August to 2021 April.
A significant ($\sim4\sigma$) hardening at $\sim$ 1 GeV is found in the $\gamma$-ray spectrum 
during a moderately flaring state  (MJD 55050-55350).
The photon index below and above the break energy is $\Gamma_1=2.36\pm0.31$ and $\Gamma_2=1.33\pm0.11$, respectively.
In the rest of the observations, the $\gamma$-ray spectrum can be described by a power-law form with the photon index of $\approx1.6$.
In the frame of a one-zone synchrotron self-Compton (SSC) model, the spectral hardening is interpreted as the transition between the synchrotron component and the SSC component.
This could be the result of a slight increase of the break/maximum Lorentz factor of the electrons.
\end{abstract}

\begin{keywords}
gamma-rays: galaxies -- galaxies: active -- radiation mechanisms: nonthermal
\end{keywords}



\section{Introduction}

Blazars are a subclass of active galactic nuclei (AGN) with their relativistic jets pointing toward us, 
which makes the jet emission extremely beaming. 
Blazars are divided into BL Lac Objects (BL Lacs) and flat-spectrum radio quasars (FSRQs) \citep[e.g.,][]{Urry1995PASP}. 
FSRQs show broad emission lines in their optical spectra while BL Lacs have spectra with weak or no emission lines.
Their radiation is dominated by nonthermal radiation in the jet, covering the entire spectrum band from radio to $\gamma$-ray energies \citep[e.g.,][]{Ulrich1997ARAA}. 

The spectral energy distribution (SED) of a blazar contains two characteristic peaks. 
The first peak appears from radio to X-ray energies, 
and the other one is in the X-ray to $\gamma$-ray ranges. The low-energy peak is considered to be the synchrotron radiation of high-energy electrons. 
The origin of the high-energy peak is still inconclusive \citep[e.g.,][]{2019Galax...7...20B}, and inverse-Compton (IC) scattering of high-energy electrons is considered as one of the popular process for producing the high-energy peak \citep[e.g.,][]{1967MNRAS.135..345R,1970RvMP...42..237B,Konigl1981,1992ApJ...397L...5M}. 

After the Fermi Gamma-ray Space Telescope operating in orbit, the Large Area Telescope (LAT) carried on Fermi has advanced the observations of GeV $\gamma$-ray emissions from blazars.
Generally,  GeV emissions from blazars display a power-law (PL) or log-parabola spectrum \citep[][]{2020ApJ...892..105A}.  
An interesting case is the significant break in the GeV spectrum of 3C 454.3 \citep{2009ApJ...699..817A,2010ApJ...721.1383A}.
This break occurs at 2-3 GeV, and the spectrum becomes softer above the break.
The change of photon index below and above the break can be as large as one, which cannot be explained by the cooling of the emitting electrons \citep{2009ApJ...699..817A}.
Several interesting models have been proposed to explain this spectral break \citep[e.g.,][]{2010ApJ...717L.118P,2010ApJ...714L.303F,2012ApJ...761....2H,2013ApJ...771L...4C,2014PASJ...66...92L,2021MNRAS.502.5875K}.

 Another interesting case is the concave GeV spectrum of 1ES 0502+675 \citep{Abdo:2010tl}.
 The LAT data collected from 2008 August 4 to 2009 February 1 exhibit a spectral hardening in the GeV spectrum of 1ES 0502+675 at $\sim$1 GeV.
 The photon index changes from $2.68\pm0.18$ to $1.47\pm0.10$. 
 This unusual concave structure is very rare.
 If confirmed, it could open interesting questions on the jet physics.
 Interestingly, this circumpolar blazar (1ES 0502+675) was suggested as one of the best neutrinos candidates expected to be associated with high-energy (PeV) cosmic neutrinos detected with IceCube \citep[][]{2017A&A...598A..36R}.
 
In this paper, we use the latest Fermi-LAT data to revisit the GeV $\gamma$-ray spectrum of 1ES 0502+675 at different time periods.
This paper is structured as follows: we give the procedure of data reduction and the $\gamma$-ray spectra in Section \ref{Data Analysis and Results}; 
In Section \ref{model} we show the modelling results for the SEDs; discussion and conclusions are presented in the last section.

\section{Data Analysis and Results}

\label{Data Analysis and Results} 

\subsection{Data Analysis}

The analysis of the data follows the standard criteria for the point-source analysis\footnote{\url{http://fermi.gsfc.nasa.gov/ssc/data/analysis/documentation/Pass8_usage.html}}. 
We employ the Science Tools package of version v11r05p3 available from {\it Fermi} Science Support Center\footnote{\url{https://fermi.gsfc.nasa.gov/ssc/data/analysis/software}} (FSSC).
The response function of the instrument is P8R3\_SOURCE\_V3, and the latest galaxy and isotropic diffusion models gll\_iem\_v07.fits and iso\_P8R3\_SO URCE\_V3\_v1.txt are used. 
The initial $\gamma$-ray model file is the LAT ten-year source catalog \citep[4FGL-DR2;][]{2020ApJ...892..105A}, and the position and spectral shape of all 4FGL-DR2 point sources within 20 degrees of 1ES 0502+675 are fixed.

Our analysis of 1ES 0502+675 is based on the Fermi-LAT Pass 8 data observed from August 4, 2008 to April 24, 2021 (MET 239557417-640988613), with energy range from 100 MeV to 300 GeV.
A 15 degree radius of interest (ROI) is selected with 1ES 0502+675 as the center.  
A 90-degree zenith angle cut is used for the data to avoid contamination from Earth's limb. 
The cut of ``(DATA\_QUAL>0)\&\&(LAT\_CONFIG==1)\&\&(angsep
(RA\_target, DEC\_target, RA\_sun, DEC\_sun)>15)" in the {\it Fermi} tool {\it gtmktime} is made to select good time intervals (GTIs) 
and to suppress the contamination from
the Sun’s emission by excluding times when the target is within 15
degrees of the Sun.
{\it Fermipy} \citep{2017ICRC...35..824W} is used to facilitate the analysis of data.

\subsection{Results}

Firstly we use the data and the initial $\gamma$-ray model files with parameters mentioned above to perform a standard Fermi-LAT likelihood analysis. 
The significance of the $\gamma$-ray emission is obtained by using the maximum likelihood test statistic (TS).
The TS is defined as 2log($L/L_{0}$), where $L$ is the maximum likelihood of the model with a point source at the target position, and $L_{0}$ is the maximum likelihood without the source \citep{1996ApJ...461..396M}.
The parameters for the sources within ROI of 5 degrees are set to be free in the fitting.
The spectrum of the target source is modeled by a power-law (PL) function, 
\begin{equation}
\frac{dN}{dE}(E)=N_0(\frac{E}{E_0})^{-\Gamma} \ .
\end{equation}

\begin{figure}
	\includegraphics[width=\columnwidth]{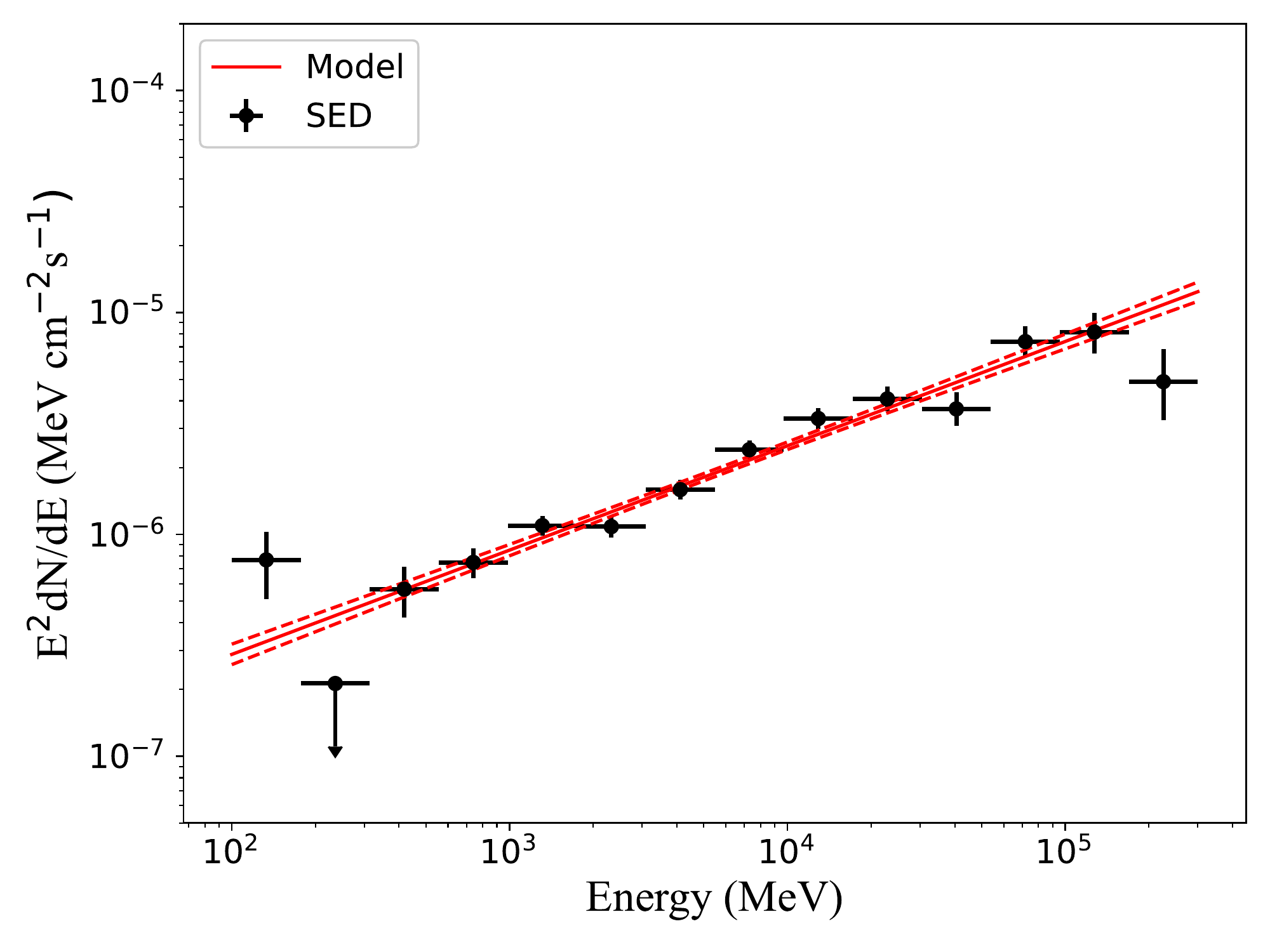}
    \caption{SED of 1ES 0502+675 during MJD 54682-59329}
    \label{fig 1}
\end{figure}

The fitting results in a total TS value of 3220. 
The average spectrum from August 4, 2008 to April 24, 2021 is showed in Figure~\ref{fig 1}.
It is described well by a single PL function with $\Gamma$=1.53$\pm$0.03.
This $\gamma$-ray spectrum is very hard, which is significantly harder than the typical $\gamma$-ray spectrum of high-synchrotron-peaked blazars (HSPs) in the Fourth Fermi Catalog of AGN \citep{2020ApJ...892..105A}\footnote{The photon index medians and rms for HSPs is $1.88\pm0.14.$}.

\begin{figure*}
	\includegraphics[width=\textwidth]{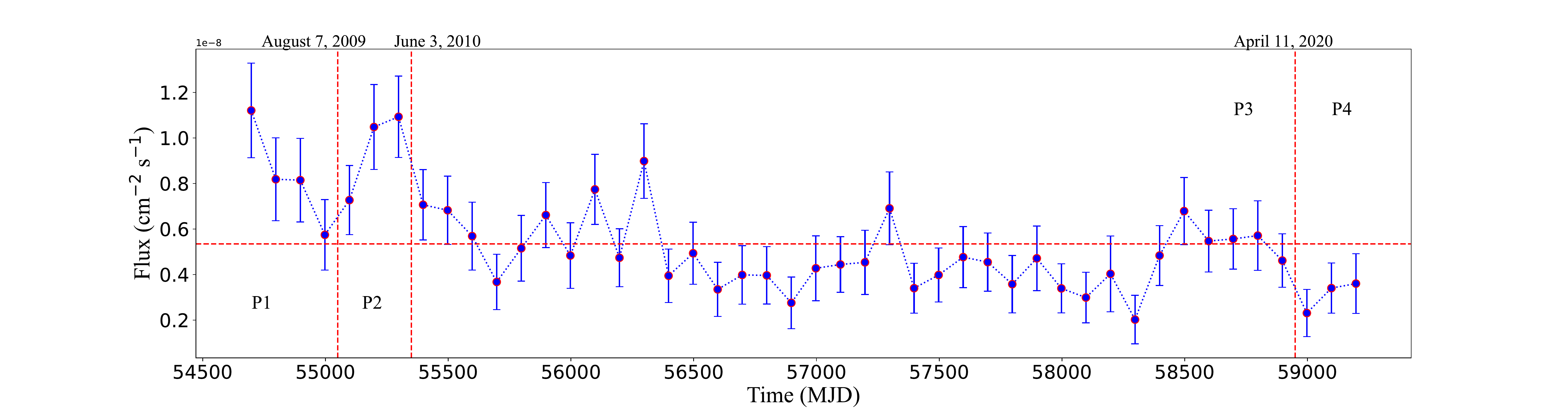}
    \caption{Light curve above 100 MeV from MJD 54682 to 59329. The horizontal dashed line denotes the average flux over the whole period ($5.4\times10^{-9}\rm \ cm^{-2}\ s^{-1}$).}
    \label{fig 2}
\end{figure*}

With the above fitting results, we generate the 100-day bin $\gamma$-ray light curve with fixed photon index and free normalization of flux for sources within 3 degrees of ROI.
For sources beyond 3 degrees of ROI, both photon index and normalization of flux are fixed.
In order to get reliable results, we exclude the time bins with the TS value less than 25.
The light curve for the whole time period is showed in Figure \ref{fig 2}.
Variation can be seen in the first two years observations.

\begin{figure*}
	\includegraphics[width=\textwidth]{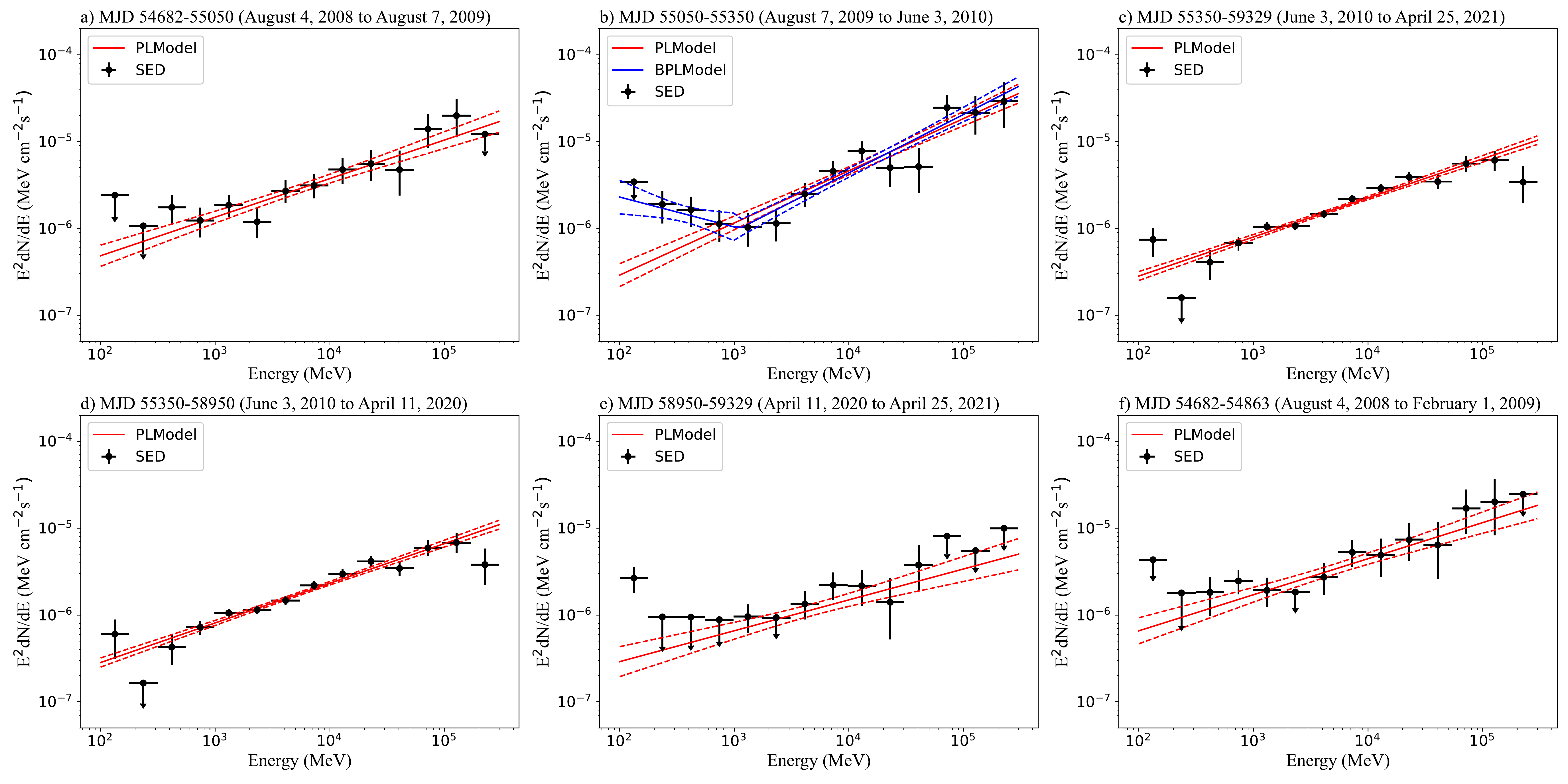}
    
    \caption{$\gamma$-ray spectra of 1ES 0502+675 in different time periods produced with the Pass 8 data.
                 LAT data are represented by black dots with error bars. The red and blue lines represent the best-fit PL and BPL models respectively. 
   }
    \label{fig_6sed}
\end{figure*}

Based on the variation showed in Figure \ref{fig 2}, we separate the entire observation into three periods: 
MJD 54682-55050 (P1), MJD 55051-55350 (P2), and MJD 55351-59329 (P3+P4).
The spectra in the three periods are produced (panels a-c in  Figure \ref{fig_6sed}).
It is found that the spectrum in P2  (panel b in  Figure \ref{fig_6sed}) deviates from a single PL form.
We then use a broken PL (BPL) function, 
\begin{equation}
	\frac{dN}{dE}(E)=\left\{
		\begin{aligned}
		&N_b(\frac{E}{E_b})^{-\Gamma_1},   & {\rm if}\   E<E_{b}\\
		&N_b(\frac{E}{E_b})^{-\Gamma_2},   & {\rm if}\   E>E_{b} 
	\end{aligned}
	\right.
\end{equation}
to perform the fitting again.
A likelihood-ratio test of the BPL and PL fit to the spectrum finds that the BPL model is better than the PL model with a test statistic
of $\rm TS_{BPL-PL}=16$, which is equal to a significance of 4$\sigma$.
From the best-fit result with the BPL model, we have $E_b=1.1\pm0.6$ (GeV), $\Gamma_1=2.36\pm0.31$, and $\Gamma_2=1.33\pm0.11$.
The spectra in P1 (panel a in Figure \ref{fig_6sed}) and P3 (panel c in Figure \ref{fig_6sed}) show a PL form with $\Gamma$=1.56$\pm$0.08 and 1.55$\pm$0.03, respectively.
Looking at the spectrum in P3, 
it is noticed that the energy flux in the first bin has an excess over the modeled PL flux.
A further analysis shows that the spectrum during MJD 55350-58950 (P3 in Figure~\ref{fig 2}) follows a PL form (panel d in Figure \ref{fig_6sed}) , 
and an excess in the first energy bin occurs after MJD 58950 (P4 in Figure~\ref{fig 2}). The spectrum in P4 is shown in panel e in Figure \ref{fig_6sed}.

We also produce the spectrum between 2008 August 4 and  2009 February 1 with the Pass 8 data (panel f in Figure~\ref{fig_6sed}).
No significant spectral hardening is found in this spectrum. 
The BPL spectrum reported by  \citet{Abdo:2010tl,2010ApJ...716...30A} cannot be confirmed.


%

    

    

\section{Modeling the SEDs}
\label{model} 

\begin{figure*}
\vspace{2.2mm} 
\centering
\includegraphics[width=0.4\textwidth] {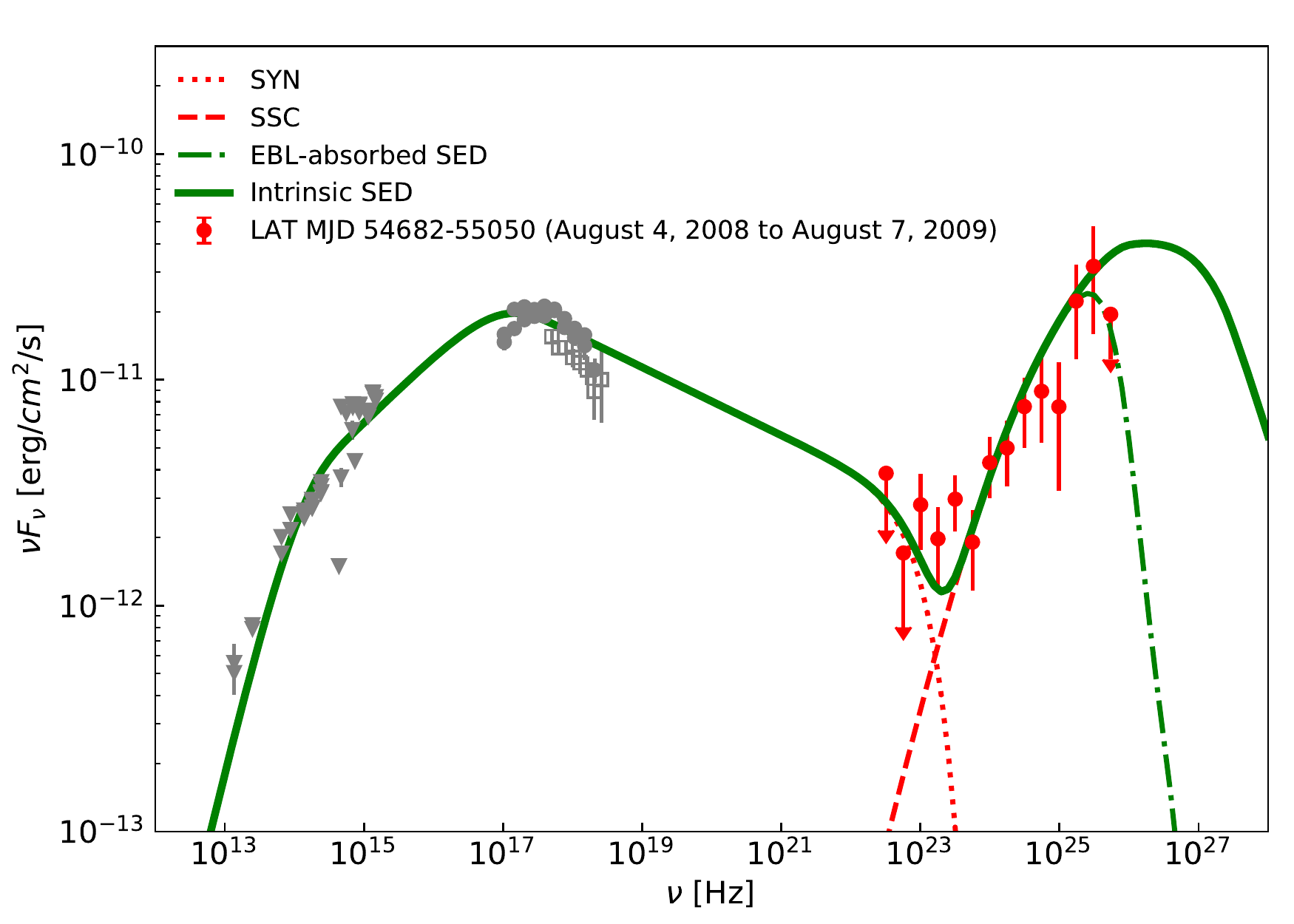}
\includegraphics[width=0.4\textwidth] {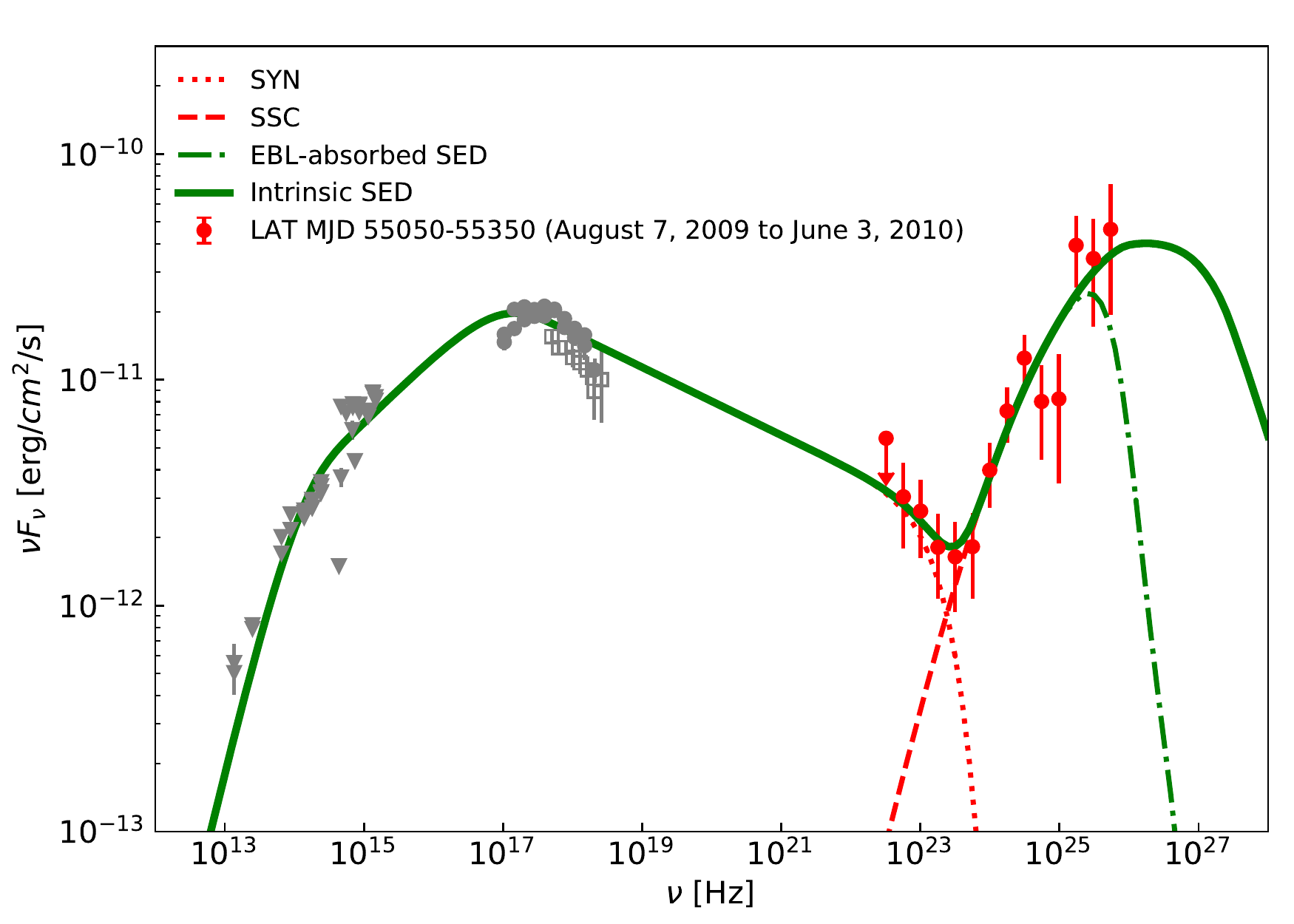}\\ 
\includegraphics[width=0.4\textwidth] {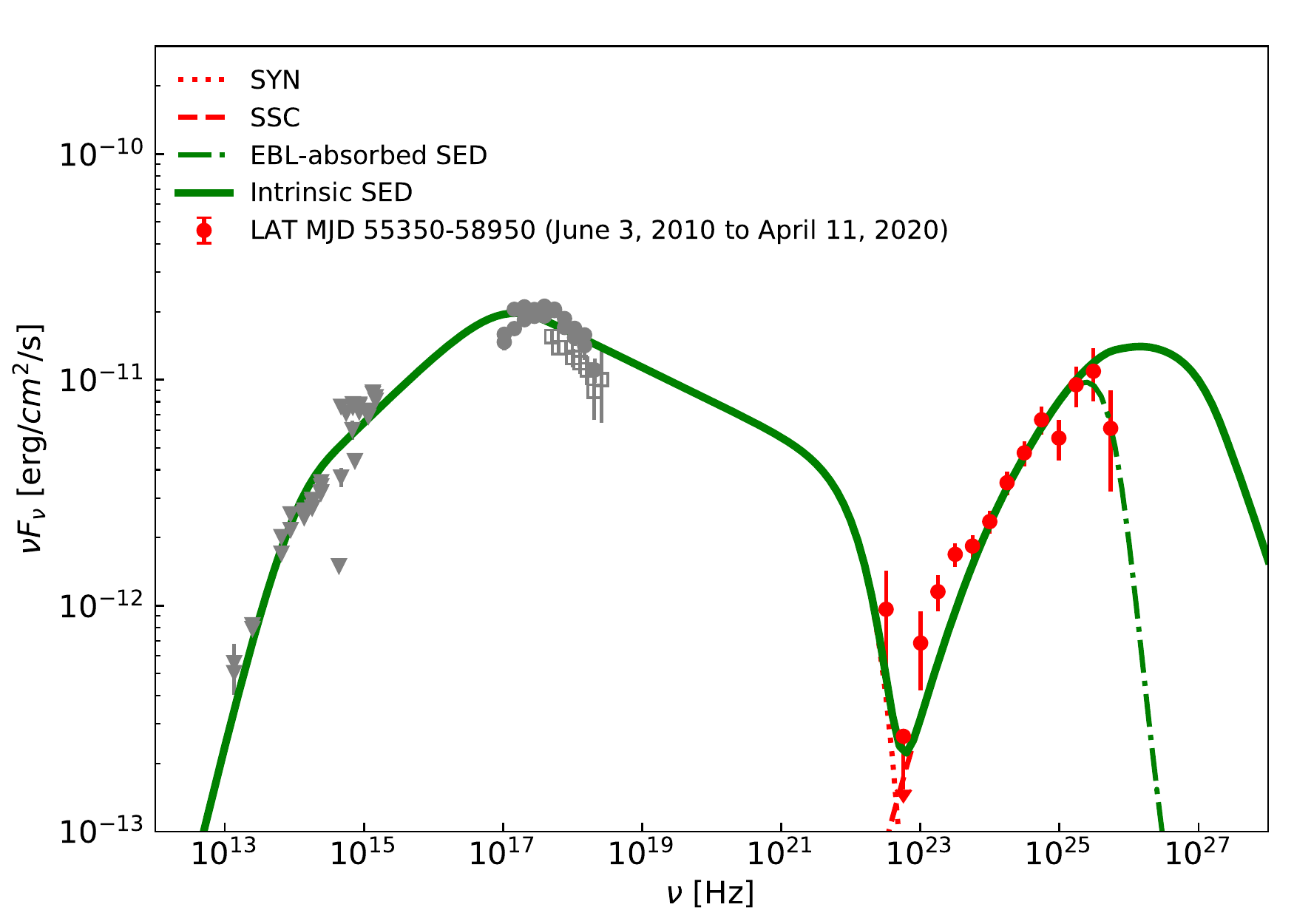} 
\includegraphics[width=0.4\textwidth] {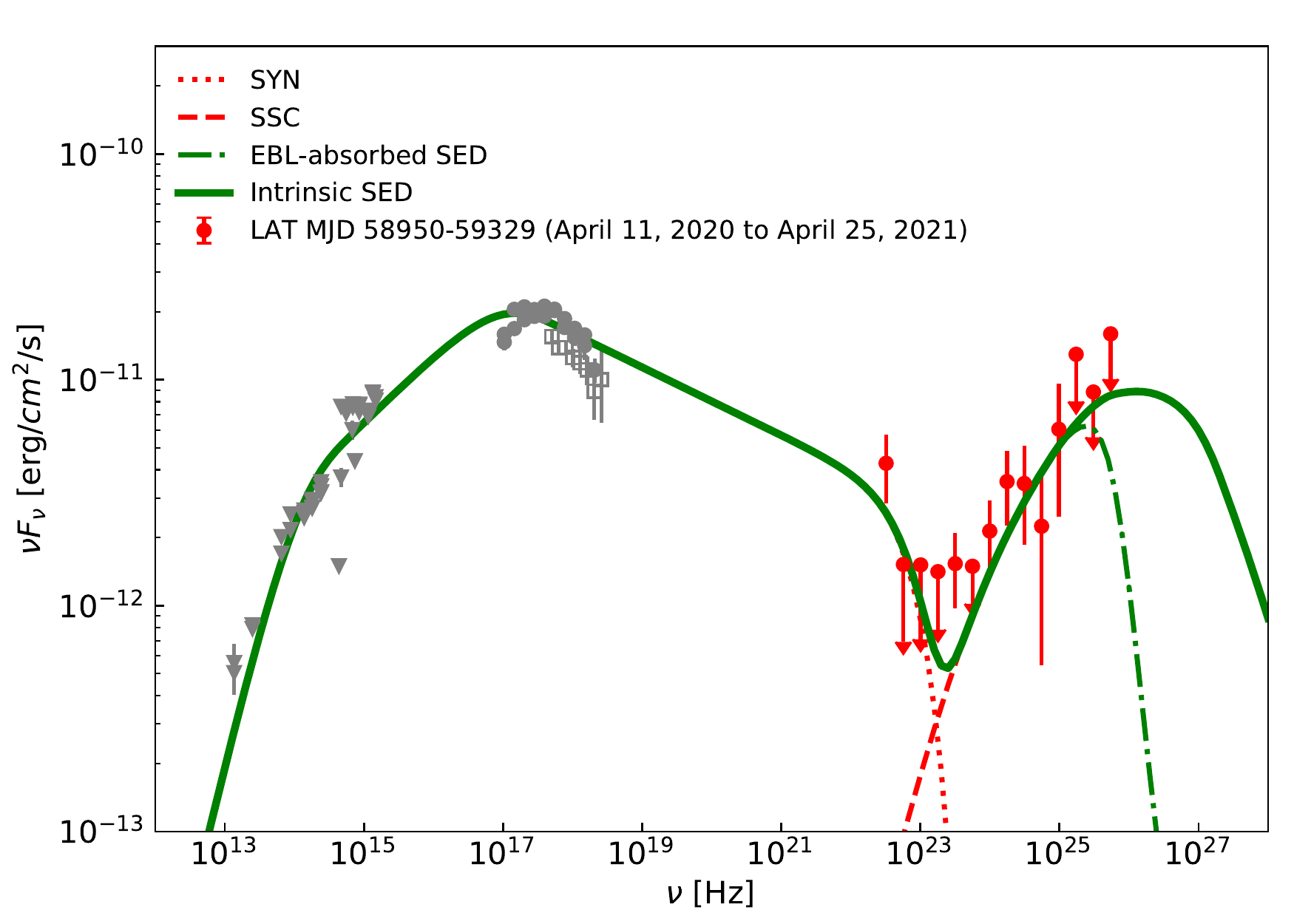} 
\caption{SSC modelling of the SEDs. The grey data are historical infrared, optical and X-ray data obtained from SSDC Sky Explorer. 
             LAT data are for the periods P1-P4.}
\label{modelb}
\vspace{2.2mm}
\end{figure*}

We collect the multi-band data through SSDC Sky Explorer\footnote{https://tools.ssdc.asi.it/}.
Broad-band  SEDs covering from infrared  wavelengths up to $\gamma$-ray energies are constructed (Figure~\ref{modelb}).
We have the X-ray data obtained from the swift-XRT observations at MJD 54833.00515 and 55155.388232 (grey filled circles).
We also show the BeppoSAX data (grey open squares) from \citet{2002babs.conf...63G}.

We adopt a one-zone synchrotron self-Compton (SSC) model to interpret the broad-band SEDs. 
In this scenario, the low-energy bump is attributed to synchrotron radiation by relativistic electrons, 
and the high-energy bump originates from IC scattering off the synchrotron photons by the same electron population.
The region responsible for the emission of 1ES 0502+675 is described as a blob of radius $R'$, 
containing a tangled magnetic field of strength $B'$, and moving towards us with a Doppler factor $\delta_{\rm D}$.
The blob is assumed to be homogeneously filled with a stationary population of electrons.
The distribution of the electrons is assumed to be a BPL which is commonly used in blazar modelling \citep[e.g.,][]{2009ApJ...692...32D,2010MNRAS.402..497G,2014MNRAS.439.2933Y},
\begin{equation}
	N_{\rm e}'(\gamma')=\left\{
		\begin{aligned}
		&N_0'(\frac{\gamma'}{\gamma'_{\rm br}})^{-n_1},   & {\rm if}\   \gamma'_{\rm min}\le\gamma'\le\gamma'_{\rm br}\\
		&N_0'(\frac{\gamma'}{\gamma'_{\rm br}})^{-n_2},   & {\rm if}\   \gamma'_{\rm br}\le\gamma'\le\gamma'_{\rm max}
	\end{aligned}
	\right.
\end{equation}
where $N_0'$ is a normalization constant; $\gamma'_{\rm min/br/max}$ is the minimum/break/maximum Lorentz factor of the electrons; and
$n_1$ and $n_2$ are the spectral indices below and above the break Lorentz factor, respectively.

The observed synchrotron spectrum (in units of $\rm erg\ cm^{-2}\ s^{-1}$) is calculated by \citep{2008ApJ...686..181F}
\begin{equation}
f_\epsilon^{\rm syn}=\frac{\delta_{\rm D}^4\sqrt{3}e^3B^\prime}{4\pi hd_{\rm L}^2}\epsilon^\prime V_{\rm b}'\int_1^\infty d\gamma^\prime N_{\rm e}^\prime(\gamma^\prime) R_{\rm s}(\epsilon^\prime/\epsilon^\prime_{\rm c}),
\end{equation}
where $e$ is the fundamental charge, $h$ is Planck's constant, $V_{\rm b}^\prime=4\pi {R^\prime}^3/3$ is the intrinsic volume of the blob, 
$d_{\rm L}$ is the luminosity distance of the source at a redshift of $z$.
Here, the function $R_{\rm s}(x)$ is the monochromatic emission power
averaged over a population of electrons with randomly distributed pitch angle \citep{1986A&A...164L..16C}, 
and an accurate approximation given by \cite{2008ApJ...686..181F} is adopted in the calculation.


The observed SSC spectrum is given by \citep[e.g.,][]{1968PhRv..167.1159J,1970RvMP...42..237B,2009ApJ...692...32D}
\begin{equation}\label{forma:ssc}
f_{\epsilon_\gamma}^{\rm ssc}=\frac{{3c\sigma_TV_b^\prime}\delta_D^4}{16\pi d_L^2}\epsilon_\gamma'^2\int_0^\infty{}d\epsilon' \frac{u_{\rm syn}'(\epsilon')}{\epsilon'^2}\int_1^\infty {}d\gamma^\prime{}\frac{N'(\gamma')}{\gamma'^2}F_{c}(x,q),
\end{equation}
where $\sigma_{\rm T}$ is the Thomson cross section, the spectral energy density of synchrotron radiation can be calculated through
$u_{\rm syn}'(\epsilon')=\frac{f_\epsilon^{\rm syn}/\epsilon' \delta_{\rm D}^4}{4\pi R^{\prime2}c}$, and
\begin{eqnarray}
  F_{\rm c}(x,q) &=& \Big[2q\ln{q}+q+1-2q^2 + \frac{(xq)^2}{2(1+xq)}(1-q)\Big] \nonumber\\
   &\times& H(q;\frac{1}{4\gamma'^2},1),
\end{eqnarray}
where
\begin{equation}
 q=\frac{\epsilon_\gamma'/\gamma'}{x(1-\epsilon_\gamma'/\gamma')},~~~~x=4\epsilon'\gamma'.
\end{equation}

Here, we use the synchrotron peak frequency $\nu_{\rm pk}$ and flux $f_{\rm pk}^{\rm syn}$ as input parameters instead of $\gamma'_{\rm br}$ and $N_0'$.
Based on the $\delta-$function approximation for synchrotron, one can obtain 
\begin{equation}
\gamma'_{\rm br}=\sqrt{\frac{\nu_{\rm pk}(1+z)}{\nu_0B'\delta_{\rm D}}}
\end{equation}
\begin{equation}
N_0'=\frac{6\pi d_L^2f_{\epsilon, \rm pk}^{\rm syn}}{c\sigma_TV_b'U_{\rm B}\delta_{\rm D}^4}\left(\frac{\nu_{\rm pk}(1+z)}{\nu_0B'\delta_{\rm D}}\right)^{-3/2}
\end{equation}
where $\nu_0\equiv4m_{\rm e}c^2/3hB_{\rm cr}$, 

The emitting region size is constrained by the relation $R'=\delta_{\rm D}t_{\rm var}c/(1+z)$, where $t_{\rm var}$ is the variability time scale in the observer frame.
Thus, the model has nine parameters: $t_{\rm var}$, $\nu_{\rm pk}$, $f_{\epsilon, \rm pk}^{\rm syn}$, $B' $, $\delta_{\rm D}$, $n_1$, $n_2$, $\gamma_{\rm min}'$ and $\gamma_{\rm max}'$. 

In the calculations, we adopt the most recent measurement of the redshift $z = 0.340$ \citep{2013ApJ...764..135S}.

From Figure~\ref{modelb}, one can see that the SEDs can be reproduced well by the SSC model.
 The intrinsic $\gamma$-ray flux is converted to the observed flux by using the extragalactic optical background light model of \citet{2010ApJ...712..238F}.
The GeV component below $\sim$1 GeV is interpreted as the tail of the synchrotron emission.
The spectral hardening during MJD 55050-55350 is caused by the transition from the synchrotron component to the SSC component.
An increase of $\gamma'_{\rm b}$ or $\gamma'_{\rm max}$ can enhance the contribution of synchrotron radiation to the observed $\gamma$-rays, 
and consequently the concave structure in the LAT spectrum will become significant.

In Table \ref{tab:table2}, we list the model parameters that are used to reproduce the observed SEDs of 1ES 0502+675 in the four epochs.
In order to reproduce the hard $\gamma$-ray spectrum, an extremely large $\gamma'_{\rm min}$ and a very small $B'$ is required, 
which is similar to the situation of the hard-TeV BL Lacs \citep[e.g.,][]{2011MNRAS.414.3566T,2018MNRAS.477.4257C}.

\begin{table*}
	\centering
	\caption{Values of the model parameters.}
	\label{tab:table2}
	\begin{tabular}{lcccc} 
		\hline
		Model 			  	 & P1 &   P2 &  P3 & P4 \\ 
		\hline
		$\rm t_{\rm var}\ $ (days)	  	 & 60				 	& 60					&60					&$60$				\\
		$\nu_{\rm pk}\ \rm (Hz)$ 	 		 & $10^{17}$		      	& $10^{17}$			&$10^{17}$			&$10^{17}$			\\
		$f_{\epsilon, \rm pk}^{\rm syn}$  	 & $2.3\times10^{-11}$	&$2.3\times10^{-11} $	&$2.3\times10^{-11}$	& $2.3\times10^{-11}$	\\
		$B' \ $ (G)			  	 & $5.0\times10^{-4}$ 	& $5.0\times10^{-4}$		&$9.5\times10^{-4}$		&$1.2\times10^{-3}$		\\
		$\delta_{\rm D}$ 		 & $17.5$			 	& $17.5$				&$17.5$				&$17.5$				\\
		$n_1$ 	 			 & $2.4$				&$2.4$				&$2.4$				&$2.4$				\\
		$n_2$ 				 & $3.3$				&$3.3$				&$3.3$ 				&$3.3$ 				\\
		$\gamma_{\rm min}'$ 	 &$6.0\times10^4$   		&$6.0\times10^4$		&$4.0\times10^4$ 		&$4.0\times10^4$ 		\\
		$\gamma_{\rm max}'$ 	 &$1.8\times10^9$   		&$2.5\times10^9$		&$5.0\times10^8$ 		&$1.0\times10^9$ 		\\
		\hline
		$R'$ (cm) 		    			& $2.03\times10^{18}$ 	&$2.03\times10^{18}$	&$2.03\times10^{18}$ 	&$2.03\times10^{18}$ 	\\
		$\gamma_{\rm br}'$ 		& $2.02\times10^6$ 		&$2.02\times10^6$ 		&$1.47\times10^6$      	&$1.31\times10^6$      	\\
		$N_{\rm 0}'\ \rm(cm^{-3})$ 			& $2.47\times10^{-12}$ 	&$2.47\times10^{-12}$      &$1.79\times10^{-12}$	&$1.59\times10^{-12}$	\\	
		\hline
	\end{tabular}
\end{table*}

\section{DISCUSSION AND CONCLUSIONS}

The early Fermi-LAT observations (the first five months) of 1ES 0502+675 showed a spectral hardening at $\sim$ 1 GeV in its LAT spectrum \citep{Abdo:2010tl,2010ApJ...716...30A}.
If confirmed, this could be an interesting spectral feature.
We perform the analysis with the Fermi Pass 8 data in the energy range between 100 MeV and 300 GeV.
Our results show that no significant spectral hardening is found in its $\gamma$-ray spectrum produced with the first seven months observations.
Furthermore, a clear spectral hardening is found in a relative high state from 2009 August 7 to 2010 June 3.
The spectrum becomes harder at $\sim$ 1 GeV with the photon index varying from 2.4 to 1.3.
For this spectrum, a BPL model is preferred over a PL model with a significance of $4\sigma$.

This kind of spectrum is usually considered as evidence of two-component model \citep[e.g.,][]{Abdo:2010tl,2010ApJ...716...30A,2012A&A...537A..47K}.
The disadvantage of the two-component model is that there are a large number of free parameters.
Intensive multiwavelengths observations are required to constrain model parameters \citep[e.g.,][]{2020A&A...640A.132M}.
Here, we interpret the SED with a one-zone SSC model. The historical infrared, optical and X-ray data are used to put a general constraint on our model.
In this model, the $\gamma$-ray emission below 1 GeV is the tail of the synchrotron component, and the emission above 1 GeV is produced by SSC process.
An increase of the $\gamma'_{\rm br/max}$ will produce such a result.
The excess in the first energy bin occurs after MJD 58950 indicates a slight increase of the $\gamma'_{\rm br/max}$.

The spectrum above 1 GeV of 1ES 0502+675  is very hard, similar to the GeV spectra of the hard-TeV BL Lacs.
To produce such a hard spectrum, the extreme model parameters ($B'$ and $\gamma'_{\rm min}$) are needed, 
which is also found in previous works \citep[e.g.,][]{2011ApJ...740...64L,2011MNRAS.414.3566T,2018MNRAS.477.4257C}.

Recently, \citet{2021ApJ...915...59Z} fitted the average SED of 1ES 0502+675 with a one-zone SSC mode by assuming a log-parabolic electron distribution.
The $\gamma$-ray data considered in \citet{2021ApJ...915...59Z} are the results in \citet[][]{2012ApJS..199...31N} and \citet[][]{2016ApJS..222....5A}, 
which cover the energy range between 2 GeV and 1 TeV.
The values of $B'$ and $\delta_{\rm D}$ derived by \citet{2021ApJ...915...59Z} are close to the values we obtained in P1.

\citet{2017PhRvD..95f3018B} reported a significant ($>5\sigma$) hardening in the Fermi-LAT spectrum of the radio galaxy Centaurus A (Cen A).
This spectral hardening occurs at $\sim$2.6 GeV, and the photon index varies from $\sim$2.7 to 2.3.
The change of the photon index below and above the break energy is smaller than that of 1ES 0502+675.
The smaller break energy in the spectrum of 1ES 0502+675 prevents us obtaining a higher significance for the spectral hardening.
As far as we know, the spectral hardening of 1ES 0502+675 is the first case for blazars and the second case for AGNs.

\section*{Acknowledgements}
This paper makes use of publicly available Fermi-LAT data provided online by the NASA-GSFC Fermi Science Support Center. 
Part of this work is based on archival data and online services provided by the ASI Space Science Data Center.
We are grateful to the anonymous referee for useful comments which have improved this work.
We acknowledge the National Natural Science Foundation of China (NSFC-11803081). 
The work of D. H. Yan is also supported by the CAS Youth Innovation Promotion Association and Basic research Program of Yunnan Province (202001AW070013).

\section*{data availability}
Data available on request.



\bibliographystyle{mnras}
\bibliography{sample} 

\begin{thebibliography}{}
\makeatletter
\relax
\def\mn@urlcharsother{\let\do\@makeother \do\$\do\&\do\#\do\^\do\_\do\%\do\~}
\def\mn@doi{\begingroup\mn@urlcharsother \@ifnextchar [ {\mn@doi@}
  {\mn@doi@[]}}
\def\mn@doi@[#1]#2{\def\@tempa{#1}\ifx\@tempa\@empty \href
  {http://dx.doi.org/#2} {doi:#2}\else \href {http://dx.doi.org/#2} {#1}\fi
  \endgroup}
\def\mn@eprint#1#2{\mn@eprint@#1:#2::\@nil}
\def\mn@eprint@arXiv#1{\href {http://arxiv.org/abs/#1} {{\tt arXiv:#1}}}
\def\mn@eprint@dblp#1{\href {http://dblp.uni-trier.de/rec/bibtex/#1.xml}
  {dblp:#1}}
\def\mn@eprint@#1:#2:#3:#4\@nil{\def\@tempa {#1}\def\@tempb {#2}\def\@tempc
  {#3}\ifx \@tempc \@empty \let \@tempc \@tempb \let \@tempb \@tempa \fi \ifx
  \@tempb \@empty \def\@tempb {arXiv}\fi \@ifundefined
  {mn@eprint@\@tempb}{\@tempb:\@tempc}{\expandafter \expandafter \csname
  mn@eprint@\@tempb\endcsname \expandafter{\@tempc}}}

\bibitem[\protect\citeauthoryear{{Abdo} et~al.,}{{Abdo}
  et~al.}{2009}]{2009ApJ...699..817A}
{Abdo} A.~A.,  et~al., 2009, \mn@doi [\apj] {10.1088/0004-637X/699/1/817},
  \href {https://ui.adsabs.harvard.edu/abs/2009ApJ...699..817A} {699, 817}

\bibitem[\protect\citeauthoryear{{Abdo} et~al.,}{{Abdo}
  et~al.}{2010a}]{Abdo:2010tl}
{Abdo} A.~A.,  et~al., 2010a, \mn@doi [\apj] {10.1088/0004-637X/710/2/1271},
  \href {https://ui.adsabs.harvard.edu/abs/2010ApJ...710.1271A} {710, 1271}

\bibitem[\protect\citeauthoryear{{Abdo} et~al.,}{{Abdo}
  et~al.}{2010b}]{2010ApJ...716...30A}
{Abdo} A.~A.,  et~al., 2010b, \mn@doi [\apj] {10.1088/0004-637X/716/1/30},
  \href {https://ui.adsabs.harvard.edu/abs/2010ApJ...716...30A} {716, 30}

\bibitem[\protect\citeauthoryear{{Acciari} et~al.,}{{Acciari}
  et~al.}{2020}]{2020A&A...640A.132M}
{Acciari} V.~A.,  et~al., 2020, \mn@doi [\aap] {10.1051/0004-6361/202037811},
  \href {https://ui.adsabs.harvard.edu/abs/2020A&A...640A.132M} {640, A132}

\bibitem[\protect\citeauthoryear{{Ackermann} et~al.,}{{Ackermann}
  et~al.}{2010}]{2010ApJ...721.1383A}
{Ackermann} M.,  et~al., 2010, \mn@doi [\apj] {10.1088/0004-637X/721/2/1383},
  \href {https://ui.adsabs.harvard.edu/abs/2010ApJ...721.1383A} {721, 1383}

\bibitem[\protect\citeauthoryear{Ackermann et al.}{2016}]{2016ApJS..222....5A} Ackermann M., Ajello M., Atwood W.~B., Baldini L., Ballet J., Barbiellini G., Bastieri D., et al., 2016, ApJS, 222, 5. doi:10.3847/0067-0049/222/1/5


\bibitem[\protect\citeauthoryear{{Ajello} et~al.,}{{Ajello}
  et~al.}{2020}]{2020ApJ...892..105A}
{Ajello} M.,  et~al., 2020, \mn@doi [\apj] {10.3847/1538-4357/ab791e}, \href
  {https://ui.adsabs.harvard.edu/abs/2020ApJ...892..105A} {892, 105}

\bibitem[\protect\citeauthoryear{{Blumenthal} \& {Gould}}{{Blumenthal} \&
  {Gould}}{1970}]{1970RvMP...42..237B}
{Blumenthal} G.~R.,  {Gould} R.~J.,  1970, \mn@doi [RvMP]
  {10.1103/RevModPhys.42.237}, \href
  {https://ui.adsabs.harvard.edu/abs/1970RvMP...42..237B} {42, 237}

\bibitem[\protect\citeauthoryear{B{\"o}ttcher}{2019}]{2019Galax...7...20B} B{\"o}ttcher M., 2019, Galax, 7, 20. doi:10.3390/galaxies7010020


\bibitem[\protect\citeauthoryear{{Brown}, {B{\r{A}}`hm}, {Graham}, {Lacroix},
  {Chadwick}  \& {Silk}}{{Brown} et~al.}{2017}]{2017PhRvD..95f3018B}
{Brown} A.~M.,  {B{\r{A}}`hm} C.,  {Graham} J.,  {Lacroix} T.,  {Chadwick} P.,
   {Silk} J.,  2017, \mn@doi [\prd] {10.1103/PhysRevD.95.063018}, \href
  {https://ui.adsabs.harvard.edu/abs/2017PhRvD..95f3018B} {95, 063018}

\bibitem[\protect\citeauthoryear{{Cerruti}, {Dermer}, {Lott}, {Boisson}  \&
  {Zech}}{{Cerruti} et~al.}{2013}]{2013ApJ...771L...4C}
{Cerruti} M.,  {Dermer} C.~D.,  {Lott} B.,  {Boisson} C.,   {Zech} A.,  2013,
  \mn@doi [\apjl] {10.1088/2041-8205/771/1/L4}, \href
  {https://ui.adsabs.harvard.edu/abs/2013ApJ...771L...4C} {771, L4}

\bibitem[\protect\citeauthoryear{{Costamante}, {Bonnoli}, {Tavecchio},
  {Ghisellini}, {Tagliaferri}  \& {Khangulyan}}{{Costamante}
  et~al.}{2018}]{2018MNRAS.477.4257C}
{Costamante} L.,  {Bonnoli} G.,  {Tavecchio} F.,  {Ghisellini} G.,
  {Tagliaferri} G.,   {Khangulyan} D.,  2018, \mn@doi [\mnras]
  {10.1093/mnras/sty857}, \href
  {https://ui.adsabs.harvard.edu/abs/2018MNRAS.477.4257C} {477, 4257}

\bibitem[\protect\citeauthoryear{{Crusius} \& {Schlickeiser}}{{Crusius} \&
  {Schlickeiser}}{1986}]{1986A&A...164L..16C}
{Crusius} A.,  {Schlickeiser} R.,  1986, \aap, \href
  {https://ui.adsabs.harvard.edu/abs/1986A&A...164L..16C} {164, L16}

\bibitem[\protect\citeauthoryear{{Dermer}, {Finke}, {Krug}  \&
  {B{\"o}ttcher}}{{Dermer} et~al.}{2009}]{2009ApJ...692...32D}
{Dermer} C.~D.,  {Finke} J.~D.,  {Krug} H.,   {B{\"o}ttcher} M.,  2009, \mn@doi
  [\apj] {10.1088/0004-637X/692/1/32}, \href
  {https://ui.adsabs.harvard.edu/abs/2009ApJ...692...32D} {692, 32}

\bibitem[\protect\citeauthoryear{{Finke} \& {Dermer}}{{Finke} \&
  {Dermer}}{2010}]{2010ApJ...714L.303F}
{Finke} J.~D.,  {Dermer} C.~D.,  2010, \mn@doi [\apjl]
  {10.1088/2041-8205/714/2/L303}, \href
  {https://ui.adsabs.harvard.edu/abs/2010ApJ...714L.303F} {714, L303}

\bibitem[\protect\citeauthoryear{{Finke}, {Dermer}  \& {B{\"o}ttcher}}{{Finke}
  et~al.}{2008}]{2008ApJ...686..181F}
{Finke} J.~D.,  {Dermer} C.~D.,   {B{\"o}ttcher} M.,  2008, \mn@doi [\apj]
  {10.1086/590900}, \href
  {https://ui.adsabs.harvard.edu/abs/2008ApJ...686..181F} {686, 181}

\bibitem[\protect\citeauthoryear{{Finke}, {Razzaque}  \& {Dermer}}{{Finke}
  et~al.}{2010}]{2010ApJ...712..238F}
{Finke} J.~D.,  {Razzaque} S.,   {Dermer} C.~D.,  2010, \mn@doi [\apj]
  {10.1088/0004-637X/712/1/238}, \href
  {https://ui.adsabs.harvard.edu/abs/2010ApJ...712..238F} {712, 238}

\bibitem[\protect\citeauthoryear{{Ghisellini}, {Tavecchio}, {Foschini},
  {Ghirlanda}, {Maraschi}  \& {Celotti}}{{Ghisellini}
  et~al.}{2010}]{2010MNRAS.402..497G}
{Ghisellini} G.,  {Tavecchio} F.,  {Foschini} L.,  {Ghirlanda} G.,  {Maraschi}
  L.,   {Celotti} A.,  2010, \mn@doi [\mnras]
  {10.1111/j.1365-2966.2009.15898.x}, \href
  {https://ui.adsabs.harvard.edu/abs/2010MNRAS.402..497G} {402, 497}

\bibitem[\protect\citeauthoryear{{Giommi}, {Capalbi}, {Fiocchi}, {Memola},
  {Perri}, {Piranomonte}, {Rebecchi}  \& {Massaro}}{{Giommi}
  et~al.}{2002}]{2002babs.conf...63G}
{Giommi} P.,  {Capalbi} M.,  {Fiocchi} M.,  {Memola} E.,  {Perri} M.,
  {Piranomonte} S.,  {Rebecchi} S.,   {Massaro} E.,  2002, in {Giommi} P.,
  {Massaro} E.,   {Palumbo} G.,  eds, Blazar Astrophysics with BeppoSAX and
  Other Observatories. p.~63 (\mn@eprint {arXiv} {astro-ph/0209596})

\bibitem[\protect\citeauthoryear{{Harris}, {Daniel}  \& {Chadwick}}{{Harris}
  et~al.}{2012}]{2012ApJ...761....2H}
{Harris} J.,  {Daniel} M.~K.,   {Chadwick} P.~M.,  2012, \mn@doi [\apj]
  {10.1088/0004-637X/761/1/2}, \href
  {https://ui.adsabs.harvard.edu/abs/2012ApJ...761....2H} {761, 2}

\bibitem[\protect\citeauthoryear{{Jones}}{{Jones}}{1968}]{1968PhRv..167.1159J}
{Jones} F.~C.,  1968, \mn@doi [PhRv] {10.1103/PhysRev.167.1159}, \href
  {https://ui.adsabs.harvard.edu/abs/1968PhRv..167.1159J} {167, 1159}

\bibitem[\protect\citeauthoryear{{Kang}, {Zheng}, {Wu}, {Chen}  \&
  {Yin}}{{Kang} et~al.}{2021}]{2021MNRAS.502.5875K}
{Kang} S.-J.,  {Zheng} Y.-G.,  {Wu} Q.,  {Chen} L.,   {Yin} Y.,  2021, \mn@doi
  [\mnras] {10.1093/mnras/stab489}, \href
  {https://ui.adsabs.harvard.edu/abs/2021MNRAS.502.5875K} {502, 5875}

\bibitem[\protect\citeauthoryear{{Katarzy{\'n}ski}}{{Katarzy{\'n}ski}}{2012}]{2012A&A...537A..47K}
{Katarzy{\'n}ski} K.,  2012, \mn@doi [\aap] {10.1051/0004-6361/201116839},
  \href {https://ui.adsabs.harvard.edu/abs/2012A&A...537A..47K} {537, A47}

\bibitem[\protect\citeauthoryear{{Konigl}}{{Konigl}}{1981}]{Konigl1981}
{Konigl} A.,  1981, \mn@doi [\apj] {10.1086/158638}, \href
  {https://ui.adsabs.harvard.edu/abs/1981ApJ...243..700K} {243, 700}

\bibitem[\protect\citeauthoryear{{Lefa}, {Rieger}  \& {Aharonian}}{{Lefa}
  et~al.}{2011}]{2011ApJ...740...64L}
{Lefa} E.,  {Rieger} F.~M.,   {Aharonian} F.,  2011, \mn@doi [\apj]
  {10.1088/0004-637X/740/2/64}, \href
  {https://ui.adsabs.harvard.edu/abs/2011ApJ...740...64L} {740, 64}

\bibitem[\protect\citeauthoryear{{Lei} \& {Wang}}{{Lei} \&
  {Wang}}{2014}]{2014PASJ...66...92L}
{Lei} M.,  {Wang} J.,  2014, \mn@doi [\pasj] {10.1093/pasj/psu067}, \href
  {https://ui.adsabs.harvard.edu/abs/2014PASJ...66...92L} {66, 92}

\bibitem[\protect\citeauthoryear{{Maraschi}, {Ghisellini}  \&
  {Celotti}}{{Maraschi} et~al.}{1992}]{1992ApJ...397L...5M}
{Maraschi} L.,  {Ghisellini} G.,   {Celotti} A.,  1992, \mn@doi [\apjl]
  {10.1086/186531}, \href
  {https://ui.adsabs.harvard.edu/abs/1992ApJ...397L...5M} {397, L5}

\bibitem[\protect\citeauthoryear{{Mattox} et~al.,}{{Mattox}
  et~al.}{1996}]{1996ApJ...461..396M}
{Mattox} J.~R.,  et~al., 1996, \mn@doi [\apj] {10.1086/177068}, \href
  {https://ui.adsabs.harvard.edu/abs/1996ApJ...461..396M} {461, 396}

\bibitem[\protect\citeauthoryear{Nolan et al.}{2012}]{2012ApJS..199...31N} Nolan P.~L., Abdo A.~A., Ackermann M., Ajello M., Allafort A., Antolini E., Atwood W.~B., et al., 2012, ApJS, 199, 31. doi:10.1088/0067-0049/199/2/31

\bibitem[\protect\citeauthoryear{{Poutanen} \& {Stern}}{{Poutanen} \&
  {Stern}}{2010}]{2010ApJ...717L.118P}
{Poutanen} J.,  {Stern} B.,  2010, \mn@doi [\apjl]
  {10.1088/2041-8205/717/2/L118}, \href
  {https://ui.adsabs.harvard.edu/abs/2010ApJ...717L.118P} {717, L118}

\bibitem[\protect\citeauthoryear{Righi, Tavecchio, \& Guetta}{2017}]{2017A&A...598A..36R} Righi C., Tavecchio F., Guetta D., 2017, A\&A, 598, A36. doi:10.1051/0004-6361/201629412

\bibitem[\protect\citeauthoryear{{Rees}}{{Rees}}{1967}]{1967MNRAS.135..345R}
{Rees} M.~J.,  1967, \mn@doi [\mnras] {10.1093/mnras/135.4.345}, \href
  {https://ui.adsabs.harvard.edu/abs/1967MNRAS.135..345R} {135, 345}

\bibitem[\protect\citeauthoryear{{Shaw} et~al.,}{{Shaw}
  et~al.}{2013}]{2013ApJ...764..135S}
{Shaw} M.~S.,  et~al., 2013, \mn@doi [\apj] {10.1088/0004-637X/764/2/135},
  \href {https://ui.adsabs.harvard.edu/abs/2013ApJ...764..135S} {764, 135}

\bibitem[\protect\citeauthoryear{{Tavecchio}, {Ghisellini}, {Bonnoli}  \&
  {Foschini}}{{Tavecchio} et~al.}{2011}]{2011MNRAS.414.3566T}
{Tavecchio} F.,  {Ghisellini} G.,  {Bonnoli} G.,   {Foschini} L.,  2011,
  \mn@doi [\mnras] {10.1111/j.1365-2966.2011.18657.x}, \href
  {https://ui.adsabs.harvard.edu/abs/2011MNRAS.414.3566T} {414, 3566}

\bibitem[\protect\citeauthoryear{{Ulrich}, {Maraschi}  \& {Urry}}{{Ulrich}
  et~al.}{1997}]{Ulrich1997ARAA}
{Ulrich} M.-H.,  {Maraschi} L.,   {Urry} C.~M.,  1997, \mn@doi [\araa]
  {10.1146/annurev.astro.35.1.445}, \href
  {https://ui.adsabs.harvard.edu/abs/1997ARA&A..35..445U} {35, 445}

\bibitem[\protect\citeauthoryear{{Urry} \& {Padovani}}{{Urry} \&
  {Padovani}}{1995}]{Urry1995PASP}
{Urry} C.~M.,  {Padovani} P.,  1995, \mn@doi [\pasp] {10.1086/133630}, \href
  {https://ui.adsabs.harvard.edu/abs/1995PASP..107..803U} {107, 803}

\bibitem[\protect\citeauthoryear{{Wood}, {Caputo}, {Charles}, {Di Mauro},
  {Magill}, {Perkins}  \& {Fermi-LAT Collaboration}}{{Wood}
  et~al.}{2017}]{2017ICRC...35..824W}
{Wood} M.,  {Caputo} R.,  {Charles} E.,  {Di Mauro} M.,  {Magill} J.,
  {Perkins} J.~S.,   {Fermi-LAT Collaboration} 2017, in 35th International
  Cosmic Ray Conference (ICRC2017). p.~824 (\mn@eprint {arXiv} {1707.09551})

\bibitem[\protect\citeauthoryear{{Yan}, {Zeng}  \& {Zhang}}{{Yan}
  et~al.}{2014}]{2014MNRAS.439.2933Y}
{Yan} D.,  {Zeng} H.,   {Zhang} L.,  2014, \mn@doi [\mnras]
  {10.1093/mnras/stu146}, \href
  {https://ui.adsabs.harvard.edu/abs/2014MNRAS.439.2933Y} {439, 2933}

\bibitem[\protect\citeauthoryear{Zhou et al.}{2021}]{2021ApJ...915...59Z} Zhou R.~X., Zheng Y.~G., Zhu K.~R., Kang S.~J., 2021, ApJ, 915, 59. doi:10.3847/1538-4357/abfe69

\makeatother
\end{thebibliography}








\bsp	
\label{lastpage}
\end{document}